\documentclass[journal]{IEEEtran}

\usepackage{titlesec}
\usepackage[utf8]{inputenc}
\usepackage{soul}
\usepackage{lettrine}
\usepackage{todonotes}
\usepackage{cite}
\usepackage{supertabular,booktabs}
\usepackage{float}
\usepackage{comment}
\usepackage{makecell}
\usepackage{graphicx}
\usepackage{pdfpages}
\usepackage{amsmath}
\usepackage{algorithm}
\usepackage{algpseudocode}
\usepackage{subfig}
\usepackage{amssymb}
\usepackage{relsize}
\newcolumntype{L}{>{\centering\arraybackslash}m{6.5cm}}
\makeatletter
\newcommand\fs@betterruled{%
  \def\@fs@cfont{\bfseries}\let\@fs@capt\floatc@ruled
  \def\@fs@pre{\vspace*{8pt}\hrule height.8pt depth0pt \kern2pt}%
  \def\@fs@post{\kern2pt\hrule\relax}%
  \def\@fs@mid{\kern2pt\hrule\kern2pt}%
  \let\@fs@iftopcapt\iftrue}
  
\floatstyle{betterruled}
\restylefloat{algorithm}

\def\BibTeX{{\rm B\kern-.05em{\sc i\kern-.025em b}\kern-.08em
    T\kern-.1667em\lower.7ex\hbox{E}\kern-.125emX}}
    
\algrenewcommand\algorithmicrequire{\textbf{Input:}}
\algrenewcommand\algorithmicensure{\textbf{Output:}}
\algnewcommand\algorithmicforeach{\textbf{for each}}
\algdef{S}[FOR]{ForEach}[1]{\algorithmicforeach\ #1\ \algorithmicdo}

\begin{document}
\title{Enhancing Mutual Trustworthiness in Federated Learning for Data-Rich Smart Cities}
\author{
    \IEEEauthorblockN{Osama Wehbi$^{1,2,3}$, Sarhad Arisdakessian$^{2,3}$, Mohsen Guizani$^{1}$, Omar Abdel Wahab$^2$, Azzam Mourad$^{4,3}$, Hadi Otrok$^{5}$, Hoda Al khzaimi$^{6}$ and Bassem Ouni$^{7}$}\\
    \IEEEauthorblockA{$^1$Mohammad Bin Zayed University of Artificial Intelligence, Abu Dhabi, UAE}\\
    \IEEEauthorblockA{$^2$Department of Computer and Software Engineering, Polytechnique Montréal, Montreal, Quebec, Canada}\\
    \IEEEauthorblockA{\normalsize$^3$Artificial Intelligence \& Cyber Systems Research Center, Department of CSM, Lebanese American University}\\
    \IEEEauthorblockA{$^4$KU 6G Research Center, Department of CS, Khalifa University, UAE}\\
     \IEEEauthorblockA{$^5$Department of CS, Center of Cyber-Physical Systems (C2PS), Khalifa University, Abu Dhabi, UAE}\\
     \IEEEauthorblockA{$^6$Division of Engineering, New York University, Abu Dhabi, UAE}\\
     \IEEEauthorblockA{$^7$AI and Digital Science Research Center, Technology Innovation Institute (TII), Abu Dhabi, UAE}\\
    \IEEEauthorblockA{{osama.wehbi}@polymtl.ca,\ {mohsen.guizani}@mbzuai.ac.ae,\ {omar.abdul-wahab, sarhad.Arisdakessian}@polymtl.ca, azzam.mourad@ku.ac.ae,\ hadi.otrok@ku.ac.ae, ha59@nyu.edu, Bassem.Ouni@tii.ae}
}

\maketitle
\begin{abstract}

Federated learning is a promising collaborative and privacy-preserving machine learning approach in data-rich smart cities. Nevertheless, the inherent heterogeneity of these urban environments presents a significant challenge in selecting trustworthy clients for collaborative model training. The usage of traditional approaches, such as the random client selection technique, poses several threats to the system's integrity due to the possibility of malicious client selection. Primarily, the existing literature focuses on assessing the trustworthiness of clients, neglecting the crucial aspect of trust in federated servers. To bridge this gap, in this work, we propose a novel framework that addresses the mutual trustworthiness in federated learning by considering the trust needs of both the client and the server. Our approach entails: (1) Creating preference functions for servers and clients, allowing them to rank each other based on trust scores, (2) Establishing a reputation-based recommendation system leveraging multiple clients to assess newly connected servers, (3) Assigning credibility scores to recommending devices for better server trustworthiness measurement, (4) Developing a trust assessment mechanism for smart devices using a statistical Interquartile Range (IQR) method, (5) Designing intelligent matching algorithms considering the preferences of both parties. Based on simulation and experimental results, our approach outperforms baseline methods by increasing trust levels, global model accuracy, and reducing non-trustworthy clients in the system.

\end{abstract}
\vspace{5pt}
\begin{IEEEkeywords}
Federated Learning, Trustworthiness, Game Theory, Smart-cities, IoT, Bootstrapping, Recommendation Systems
\end{IEEEkeywords} 

\section{Introduction}
\label{intro}
\lettrine{T}{h}e concept of smart cities is gaining popularity day by day. This is driven by the fact that urban regions now host more than half of the global population. Projections from the World Bank indicate that this urban population will reach 6 billion by 2045~\footnote{https://www.worldbank.org/en/topic/urbandevelopment/overview}. Smart cities rely on data collected from Internet of Things (IoT) sensors and Internet of Vehicles (IoV) networks~\cite{hani}. According to the International Data Corporation (IDC), the number of IoT devices globally is estimated to reach approximately 41.6 billion by 2025~\footnote{https://www.idc.com/}. This data abundance offers businesses opportunities to enhance production processes and strategic decision-making, boosting profitability~\cite{mario, rabab1}. Nonetheless, the prevailing data analysis method often involves offloading data of smart devices to edge servers for training and pattern extraction. This approach encounters scalability challenges given the exponential growth of smart city devices and data heterogeneity~\cite{adoom2,sarhad,adoom1, waz, jamal}.\\
To address these challenges, Federated Learning (FL) has emerged as a solution to communication and privacy concerns~\cite{i1,ahmad}. FL enables model training tasks to be executed locally, at the device level, and through distributed methods~\cite{mohsen}. In this process, the federated server initially creates a machine-learning model and communicates its weights to a selected group of client devices during each communication round. Clients perform model training on their local data and provide updated weights to the federated server, which aggregates them to construct a new global model. This iterative process continues until a specific number of training rounds or the desired accuracy is reached~\cite{i3}.\\
The distributed nature of data in smart cities drives the adoption of FL, aligning with data localization and privacy needs. FL handles data growth, offering scalability, resource efficiency, resilience, real-time adaptability, and latency reduction. It taps into the potential of edge devices, ensuring system resilience even in node failure scenarios. In summary, FL addresses challenges posed by distributed datasets, enhancing privacy and scalability while unlocking the potential of smart city initiatives.
Nevertheless, given the heterogeneous nature of devices in smart cities, it is a challenging task to select a trustworthy set of clients to cooperate and work within the model training during federated learning rounds. The usage of traditional approaches, such as the random client selection technique, would pose several threats to the integrity of the system. This is due to the possibility of selecting malicious clients which can target the system with attacks such as poisoning attacks. These attacks can degrade the accuracy of the final global model~\cite{guesmi2023physical}. Current literature often focuses on assessing client trust while neglecting trust in federated servers and primarily targets bootstrapping the process rather than addressing newcomer IoT device challenges~\cite{omar}.

This paper introduces a novel framework for bilateral client selection in federated learning environments in the domain of smart cities. The advantage of the approach stems from the fact that it considers the trustworthiness of both entities in the environment, i.e., the federated servers, as well as the clients. The goal is to enhance the security, reliability, and efficiency of the federated learning system while unlocking the full potential of smart cities. The main challenge lies in assessing the client's trust without directly accessing the smart devices' internals. The approach addresses this through a method based on resource monitoring and statistical IQR score analysis. We assume that by monitoring client device resource utilization, abnormal behaviour or performance deviations can raise concerns about trustworthiness. In addition, the reputation of the federated servers is also a crucial factor, which leads to the use of a reputation-based recommendation system in the environment. This system evaluates and recommends trustworthy federated servers based on past performance, integrity, and reliability while introducing a credibility score for recommending devices; by considering the trustworthiness of both clients and the federated servers, the proposed framework targets to create a robust and secure federated learning ecosystem in the realm of smart cities. This approach addresses concerns related to data privacy, security breaches, and biased learning outcomes, fostering trust and confidence among federated learning parties.

\subsection{Problem Statement}
\label{problem statement}

Performing federated learning training tasks on smart devices can provide substantial advantages. Nonetheless, adopting an unsupervised client selection approach, as per the standard federated learning model\cite{p1}, has drawbacks for a plethora of reasons. First, the inclusion of untrustworthy clients can significantly degrade the performance of the overall system. Untrustworthy devices may engage in malicious activities, such as injecting bogus data into the local training process, which can lead to skewed models and reduced accuracy~\cite{guesmi2023physical}. This not only undermines the effectiveness of the federated learning approach but also compromises the privacy and security of the system. Second, While existing research primarily focuses on assessing the trustworthiness of clients, this process is usually single-sided, as it often neglects the potential untrustworthy effects of servers, particularly in scenarios involving payment or data disclosure. In such contexts, servers can pose a threat to the confidentiality and integrity of client data. Overlooking this aspect of trust leaves the federated learning system vulnerable to data breaches and unauthorized access. To create a more comprehensive and robust trust framework, we believe that it is essential to consider the trustworthiness of both clients and servers.
Furthermore, allowing the servers to assign trust values to themselves is also a challenging task since the servers might act mischievously and assign high scores deliberately. Such factors ensure that the trustworthiness of federated servers poses a significant challenge and should not be neglected.  However, the absence of a comprehensive trust assessment framework hinders the widespread adoption of federated server infrastructures in smart cities. In our prior research~\cite{r6,r7}, we addressed the challenge of selecting clients in federated learning, taking into account client incentives (rewards) and enhancing the accuracy of the global model on federated servers. In this study, we propose that trust scores for all participants in a federated learning system should take a crucial role in the decision-making process. Consequently, we have created a mechanism to foster trust between smart devices and federated servers, which facilitates mutual trust establishment and authentication between these two parties.

\subsection{Contributions}
\label{contributions}
Motivated by the limitations of random client selection and the fact that most existing work in the current literature focuses on assessing client trust while neglecting trust in federated servers and primarily targets bootstrapping the process rather than addressing newcomer IoT device challenges. In this paper, we introduce a novel mutual trust selection approach for federated learning within smart city environments. Our approach relies on a combination of reputation-based recommendation systems, statistical IQR trust-establishing methods, and matching game theory. The reputation-based recommendation system evaluates and recommends trustworthy federated servers based on past performance, integrity, and reliability while introducing a credibility score for recommender devices. The IQR method is applied to evaluate the trustworthiness of smart devices based on resource usage. Finally, the adoption of matching game theory allows the clients and the federated servers to choose each other by preferences, taking into account trustworthiness. Consequently, smart devices will play a role in determining their preferred federated server based on the server's level of trust and vice versa.

Hereafter, we summarize the main contributions of our work:
\begin{itemize}

\item Developing a trust bootstrapping model that enables devices to allocate trust values to newcomer federated servers with no prior participation. We believe this will ensure fairness in future Federated Learning (FL) training rounds between active servers, newcomers and existing clients in terms of participant acquisition.

\item Creating a credibility score model for devices to uphold the bootstrapping quality by dynamically updating device credibility scores based on their associated trust values and recommendations provided by various recommenders (i.e., other devices in the system that can give a recommendation towards another device).

\item Introducing a novel client selection approach in federated learning, which draws inspiration from game theory and takes into consideration the preferences of both clients and servers during the selection phase.

\item Creating a different trust calculation mechanism for both client smart devices and federated servers.

\item Developing a group of matching algorithms which take into account the preferences of client devices as well as federated servers. These proposed algorithms establish a stable matching relationship, discouraging withdrawal from the matching by either clients or servers.

\end{itemize}

\subsection{Paper Organization}
\label{paperoutline}

We investigate the existing research concerning the selection of trustworthy clients, emphasizing the originality of our approach in Section \ref{RelatedWork}. We then outline the trust establishment mechanisms between smart devices and federated servers in Section \ref{problemformulation}. Subsequently, we introduce the method for calculating trust in smart devices, Section \ref{Fedtrust}, and federated servers, Section \ref{smartrust}.
Furthermore, we explore the principles of matching game theory and the formulation of preference lists in Section \ref{matchingametheoryformulation}. We present the development of distributed matching algorithms in Section \ref{matchingselection}.
To comprehensively evaluate our proposals, we describe the experimental setup employed for simulations and result interpretation in Section \ref{experiments}. Lastly, we offer a concise summary and conclusive remarks in Section \ref{Conclusion}, encapsulating the essence of our study.

\section{Related Work}
\label{RelatedWork}

In this section, we survey the existing works in the literature pertaining to client selection as well as the usage of game theory in the selection process. We also highlight the novelty of our approach.

\subsection{Client Selection in Federated Learning}
In~\cite{r1}, the authors introduce a novel method for trust formation between IoT devices and training servers. The technique focuses on identifying underused or overused resources during model training by the device by implementing a Double Deep Q Learning (DDQN) task scheduling method. This method incorporates trust scores and energy levels to perform scheduling decisions.

Chen et al. in~\cite{r2} employ an information gain strategy to assess gradient variation similarities between the global model and locally trained models. This solution operates at the model aggregation level, where trust scores are assigned to individual working devices based on similarity values. This approach mitigates the adverse effects of suspicious nodes.

In \cite{r3}, Cao et al. propose FLTrust, a federated learning approach. Where task owners bootstrap trust values for each client's model in each round using pre-obtained historical data named as root dataset. this historical data is utilized to construct a model to predict trust scores.

In this work~\cite{r4}, Wahab et al. put forth an approach that calculates trust values for potential recommender devices. This approach also incorporates a scheduling technique based on Interquartile Range statistical and double-deep Q-learning methods to specify the accepted client devices. The selection of suitable candidate devices relies on both the levels of energy and the candidate trust score.

In \cite{r5}, Bao et al. introduce FLChain, which relies on the blockchain mechanism for establishing a sustainable, distributed, and transparent federated learning scheme. This system incorporates trust factors and rewards to incentivize trustworthy trainers and penalize dishonest devices in real time.

In a recent study by authors in \cite{r8}, FLTrust was proposed to mitigate the presence of malicious clients in federated learning. Unlike conventional byzantine-robust techniques that heavily count on statistical analysis to identify malicious clients, their method leverages a small quantity of acquired data to bootstrap trust.

The work presented in \cite{nishio2019client} tackles the issue of restricted computational capabilities on client devices and introduces an approach referred to as FedCS. This method aims to address the challenge posed by the diversity of resources available on client devices by introducing certain limitations on the acceptance criteria for updated models. In \cite{abdulrahman2020fedmccs}, the authors introduce FedMCCS. This approach considers the resources of each individual device during the selection process. An evaluation of these resources is performed to determine whether a particular device is capable of performing a Federated Learning (FL) task.

Another noteworthy contribution in the field of FL is presented in \cite{r9}, the authors proposed FLOD, an innovative approach designed to defend against byzantine threats in federated learning. FLOD leverages trust establishment, aggregation based on Hamming distance, several optimizations and homomorphic encryption to guarantee resistance to byzantine attacks and privacy preservation. In contrast to other methods, these techniques illustrate the benefits of utilizing trust establishment to assign a trust score to each local model.

\subsection{Game Theory in Federated Learning}

Game theory enhances decision-making by providing a framework to analyze strategic interactions, leading to optimal outcomes in diverse fields such as economics, politics, and evolutionary biology. It promotes cooperation, reveals hidden incentives, and guides the formulation of effective strategies. 

The authors of\cite{r11} introduce a novel matching-theoretic strategy for addressing the challenge of low-latency task scheduling in multi-access edge computing networks characterized by incomplete preference lists. In a large ecosystem, the matching process takes place between the server responsible for federated learning tasks and end-user devices.

In \cite{r12}, the authors proposed a matching-theoretic solution with incomplete preference lists in a multi-access edge computing environment to handle the low-latency task scheduling issue. The matching process takes place between edge nodes responsible for federated learning tasks and end-user devices in a large environment.  

In \cite{r13}, the authors present a one-to-one modified dual-sided matching framework to handle task allocation in federated learning while safeguarding against untrustworthy clients. The study also introduces a worker reputation metric. The proposed approach effectively allocates tasks and mitigates potential malicious actions.

Based on the literature, the majority of the approaches consider trust from a single-side approach. That is, from the side of the clients or vice versa.  
Also, it is important to note that these solutions primarily focus on bootstrapping the overall model and the federated learning process rather than specifically targeting newcomer IoT devices, which is the primary focus of our work. On the other hand, our suggested approach aims to develop a mutual trust-based solution that considers the trust of both entities, i.e., the server as well as the clients. This gives room for less biased selections while minimizing the levels of untrustworthiness in the environment, making the system more robust and secure.

\section{Problem Formulation}
\label{problemformulation}
This section outlines the procedure for establishing trust scores among participants in the federated learning setting, specifically, the clients and servers. Additionally, we emphasize and provide explanations for the various symbols employed in this study, as represented in Table~\ref{table_1}.
\begin{table}[!ht]
\caption{Notations Index}
\begin{center}
\begin{tabular}{|c L|}
\hline 
 & \\
Notation & Description \\ [0.5ex] 
\hline\hline
 & \\
 & \textbf{Federated Servers  and Clients}\\
$A$ & Initial group of federated servers\\
$D$ & Initial group of clients \\
$a$ & a single federated server \\
$d$ & a single client \\
$S_a$ & Set size of clients selected by server $a$\\
$L_{S_a}$ & List of clients selected by server $a$\\
$K_a$ & Set of clients desired by server $a$\\
& \\
& \textbf{Trust Establishment}\\
$T_a$ & Trust score for a single federated server $a$\\
$T_d$ & Trust score for a single client $d$\\
$F$ & Features of resource data, namely, F = \{ RAM , CPU, bandwidth\}\\
$f$ & a single feature $f \in F$\\
$M^f_{d}$ & List of devices resource feature $f \in F$ upon a certain round $r$\\
$E^f$ & List of resource utilization by a subset $D^\prime \subset D$ of clients during training\\ 
$\alpha^f_{d}$ & Total overutilization of feature $f$ by client $d$\\
$\beta^f_{d}$ & Total underutilization of feature $f$ by client $d$\\
$\gamma^f_{d}$ & Frequency of overutilization of feature $f$ by client $d$\\
$\delta^f_{d}$ & Frequency of underutilization of feature $f$ by client $d$\\
$\zeta^f_{d}$ & Average underutilization of feature $f$ by client $d$\\
$\varepsilon^f_{d}$ & Average overutilization of feature $f$ by client $d$\\
$\eta^f_{d}$ & overutilization of feature $f$ by client $d$ according to upper limit\\
$\psi^f_{d}$ & overutilization frequency of features $f$ for device $d$ \\
$\vartheta^f_{d}$ & Underutilization of feature $f$ by client $d$ according to lower limit\\
$\lambda^f_{d}$ & Underutilization frequency of feature $f$ for client $d$ \\
& \\
& \textbf{Matching Game Approach}\\
$\Gamma(d)$ & A matching game strategy of client $d$ \\
$\Gamma$ & A matching game relation between two parties\\
$P_a$ & Preference array of server $a$\\
$P_d$ & Preference array of device $d$\\
$d^\prime \succ_s d^{\prime\prime}$ & Server $a$ prefers $d^\prime$ over $d^{\prime\prime}$\\
$\Gamma(a)$ & A matching game strategy of server $a$\\
&\\
\hline
\end{tabular}
\end{center}
\label{table_1}
\end{table}

\subsection{Trust Establishment Mechanism for Clients}
\label{Fedtrust}
In this section, we introduce the trust establishment mechanism applied by the servers to compute the client's trust score. By monitoring the clients' resources (i.e., CPU, RAM, and Bandwidth) consumption during the training rounds, the federated server will be able to specify untrustworthy clients that over-use or under-use their resources.

\begin{algorithm}[ht]
\caption{Client Devices Trust Establishment}
\label{client-pref}
\begin{algorithmic}[1]
\Require {F, $E^f$, $M^f_{d}$}
\Ensure {Trustworthiness of client device d}
\ForEach {$f$ \textbf{in} $F$ \textbf{and} $d$ \textbf{in}  $D$}
\State Arrange the data points in sample $E^f$
\State Obtain $Q^f_{3}$, $Q^f_{2}$, $Q^f_{1}$
\State Obtain the interquartile range: $IQR^f$ = $Q^f_{3}$ - $Q^f_{1}$
\State Obtain lower limit: $L^f$ = $Q^f_{1}$ + $IQR^f$ * 1.5
\State Obtain upper limit: $U^f$ = $Q^f_{3}$ + $IQR^f$ * 1.5
\ForEach {$x^f_{d}$ \textbf{in} $M^f_{d}$}
\If {$x^f_{d} > U^f$}
\State $\alpha^f_{d}$ +=$x^f_{d}$
\State $\gamma^f_{d}$ +=1
\ElsIf {$x^f_{d} < L^f$}
\State $\beta^f_{d}$ +=$x^f_{d}$
\State $\delta^f_{d}$ +=1
\EndIf
\EndFor
\If {$\gamma^f_{d} > 0$}
\State $\varepsilon^f_{d}$ = $\alpha^f_{d}$ / $\gamma^f_{d}$
\State $\eta^f_{d}$ = $U^f$ / $\varepsilon^f_{d}$
\State $\mid \psi_d\mid$ += 1
\EndIf
\If {$\delta^f_{d}  > 0$}
\State $\vartheta^f_{d}$ = $L^f$ / $\vartheta^f_{d}$
\State $\zeta^f_{d}$ = $\beta^f_{d}$ / $\delta^f_{d}$
\State $\mid \lambda_d\mid$ += 1
\EndIf
\EndFor
\If {$\mid \lambda_d\mid$ ==0 \textbf{and} $\mid \psi_d\mid$  ==0}
\State $T_d$ = 0
\Else
\State {$T_d$ = $\frac{\sum_{f}^{F} \eta^f_{d} + \vartheta^f_{d}}{\psi_d + \lambda_d}$}
\EndIf
\end{algorithmic}
\end{algorithm}

The methodology applied relies on the Interquartile Range (IQR) statistical technique~\cite{r4}. The core concept behind the IQR method involves dividing the dataset into different quartiles, denoted as Q3, Q2, and Q1. Q1 represents the threshold above which 75\% of the dataset observations lie, while Q2 corresponds to the median of the sorted dataset. Q3, on the other hand, signifies the point below which 75\% of the dataset resides. This approach takes a small dataset as input, containing the average consumption of RAM, CPU, and bandwidth across a group of clients previously involved in federated learning model training with similar resource requirements. To identify potentially suspicious usage patterns within the dataset, the federated server incorporates the Interquartile Range (IQR) statistical technique.

\subsection{Bootstrapping Trust for federated servers}
\label{smartrust}
In this section, we discuss the proposed trust bootstrapping solution and highlight the main phases of the process.

Bootstrapping is a mechanism commonly applied in various areas of study (like cloud computing) to provide recommendations to the different parties involved in roles within the environment. In the context of our work, such a system helps determine the level of trust for newly joined federated servers in the environment when there is no prior information about them~\cite{pf1, wahab2016towards}. Accordingly, our approach aims to use bootstrapping to assess the trustworthiness of newly added federated servers initially. In our proposed solution, clients create preference lists which contain a list of the federated servers they want to work with based on the trust levels of the servers. Thus, servers without trust scores are excluded from the list, as assigning random trust values can create unfair situations for both the clients and servers.  To overcome this problem, our method relies on the collaboration of multiple active clients working together in a distributed manner. Whenever a client encounters a new federated server, it seeks guidance from its neighbouring devices to obtain a recommended trust value.

\subsubsection{Bootstrapping Overview}
\label{sec:Bootstrapping Overview}

\begin{figure}[h!]
    \centering
    \includegraphics[width=0.8\linewidth]{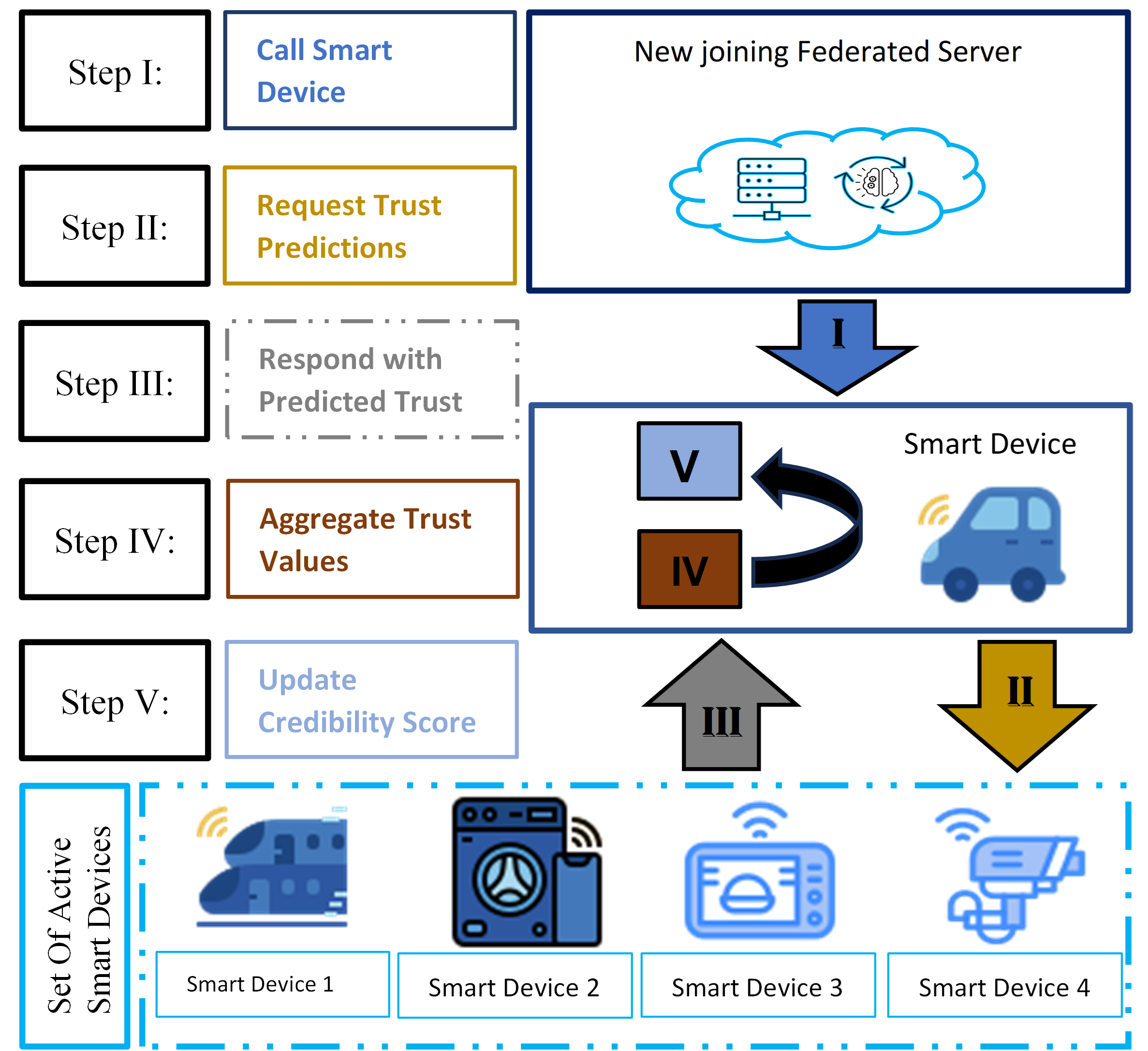}
    \caption{Bootstrapping Architecture}
    \label{fig:boot}
\end{figure}

In our collaborative smart devices ecosystem, depicted in Fig.~\ref{fig:boot}, the smart devices work together to establish a trust score for assessing the trustworthiness of federated servers. Each smart device maintains a comprehensive dataset that captures the historical interactions with the federated server in the previous rounds of training. This dataset encompasses various characteristics, such as type, region, payment, and the corresponding trust score. When a new federated server expresses its willingness to call a specific client (Step I), the client requests from the neighbouring clients to predict the expected trust score for the federated server (Step II). To accomplish this, the client forwards a bootstrapping query to all other active nearby clients (Step II). Subsequently, the intended clients predict the trust score based on their local decision tree classification model (section \ref{decision_tree}) (Step III). Once the trust values are produced, the client aggregates the received scores using the Dempster-Shafer method (DST) by considering the credibility score of each smart device (Step IV). Finally, the smart device updates the credibility score for all participated neighbours based on the DST result (Step V). This collaborative process harnesses the collective knowledge and experiences of the clients in the system to assess the federated servers' trustworthiness and integration capabilities within the smart cities ecosystem.

\subsubsection{Decision Tree Creation}
\label{decision_tree}

The decision tree is a popular supervised technique in machine learning that utilizes the concept of information gain to construct decision trees in a top-down manner. This approach, inspired by the ID3 algorithm and employing a greedy search strategy without backtracking \cite{quinlan1986induction}, finds broad application in various domains. Decision trees can be created for regression and classification tasks, where the focus of classification decision trees lies in categorical outcomes. When constructing a classification decision tree, the primary consideration is information gain. Information gain quantifies the decrease in entropy or impurity within a dataset after dividing it based on a specific attribute. The objective is to identify the attribute that maximizes information gain, leading to informative and predictive splits. As the decision tree continues to split the dataset into subsets that are increasingly homogeneous with regards to the target variable, a tree structure gradually emerges.

The decision tree begins with a root node and progressively splits the data into branches according to attribute values. At each decision node, the attribute that maximizes information gain is selected for further partitioning. Leaf nodes represent final classification outcomes or class labels. The resulting structure of nodes and branches forms a set of rules that facilitate the classification of new instances. 

Information gain is a guiding principle for determining the most informative attribute to split on at each node. It quantifies the level of uncertainty or randomness in the dataset attributable to a specific attribute. The attribute with the highest information gain is deemed the most significant in discriminating between different classes or categories. Through the iterative application of this process, the classification decision tree continuously refines its predictive capabilities, enabling accurate classification of previously unseen data. 

Below, we highlight the equations for entropy and information gain:

Entropy:
\begin{equation}
\label{eqEntropy}
H(X) = -\sum_{i=1}^{n} P(x_i) \log_2(P(x_i))
\end{equation}

Information Gain:
\begin{equation}
\label{eqInfGain}
\text{Gain}(X, A) = H(X) - \sum_{v \in \text{Values}(A)} \frac{|X_v|}{|X|} H(X_v)
\end{equation}

Where:
\begin{itemize}
\item[-] \(H(X)\) denotes the entropy of the dataset \(X\)
\item[-] \(P(x_i)\) represents the probability of occurrence of class \(x_i\) in dataset \(X\)
\item[-] \(A\) is an attribute used for splitting
\item[-] \(X_v\) denotes the subset of dataset \(X\) where attribute \(A\) takes value \(v\) 
\item[-] \(\text{Values}(A)\) represents the possible values of attribute \(A\) 
\item[-] \(|X|\) and \(|X_v|\) denote the cardinality of dataset \(X\) and subset \(X_v\) respectively

\end{itemize}

These equations capture the essence of entropy and information gain, which are fundamental concepts used in decision tree algorithms to determine the quality of attribute splits and construct accurate decision trees.
            
\bgroup
\def\arraystretch{1.5}%
\begin{table}[h!]
\caption{Historical Data Sample Example}
\begin{center}
\begin{tabular}{||c|c|c|c||}
\hline
\textbf{Server} & \textbf{Geographic Location} & \textbf{Trust Score} & \textbf{Trust Status} \\ [0.5ex]\hline
S1       & Asia    & 99.05 & YES       \\ \hline
S1       & America & 100 & YES       \\ \hline
S2       & Africa  & 99.37 & YES       \\ \hline
S2       & Africa  & 99.88 & YES       \\ \hline
S3       & America & 99.54 & NO       \\ \hline
S4       & Asia    & 73.69 & NO      \\ \hline
S4       & America & 97.62 & YES      \\ \hline
S4       & America & 92.42 & YES      \\ \hline
S4       & Africa & 87.62 & NO      \\ \hline
S4       & Europe & 82.42 & NO      \\ \hline
S4       & Asia    & -    & ?      \\ \hline
\end{tabular}
\end{center}
\label{tab:TABLE II}
\end{table}
\egroup

In order to showcase the practical application of information gain in federated learning scenarios for decision tree generation, we present a demonstrative example using a subset of the dataset presented in Table~\ref{tab:TABLE II}. The dataset comprises data collected from IoT devices about federated servers, including details such as server type, provider, deployment region, and observed trust.

For example, to assess the trustworthiness of a newly deployed federated server, S4, which is located in Asia, we initiate the evaluation process by examining the (sub)dataset. The first step involves determining the dataset's entropy using Equation (1). By examining the trust status class labels in Table 1, we observe that among the ten servers, four are classified as "No" and six as "Yes". Accordingly, the dataset's entropy can be computed as follows: \\

$E(DATA)$ = - (6/10) log2(6/10) - (4/10) log2(4/10) = 0.971 \\

Afterward, we move on to calculate the entropy of each attribute, specifically the server and its geographic location, using Equation (1). It is crucial to mention that we do not include the trust score attribute in these computations as it is unknown for newly joined servers. Nevertheless, we incorporate this attribute in Table 2 to illustrate the decision-making process for the class label attribute, trust status. 

Hereafter, we illustrate the methodology employed to calculate the entropy value for the Server attribute using equation \ref{eqEntropy}.

Server = S1: We have two instances labeled "Yes" and 0 instances labeled "No":
\begin{equation*}
E(D_{\text{S1}}) = - 1 \cdot \log_2(1) - 0 \cdot \log_2(0) = 0
\end{equation*}

Server = S2: We have two instances labeled "Yes" and 0 instances labeled "No":
\begin{equation*}
E(D_{\text{S2}}) = - 1 \cdot \log_2(1) - 0 \cdot \log_2(0) = 0
\end{equation*}

Server = S3: We have 1 instance labeled "No" and 0 instances labeled "Yes":
\begin{equation*}
E(D_{\text{S3}}) = 0
\end{equation*}

Server = S4: We have two instances labeled "Yes" and three instances labeled "No":
\begin{equation*}
E(D_{\text{S4}}) = - \left(\frac{3}{5}\right) \cdot \log_2\left(\frac{3}{5}\right) - \left(\frac{2}{5}\right) \cdot \log_2\left(\frac{2}{5}\right) = 0.971
\end{equation*}

Now, Since we have the entropy, we can move forward and calculate the information gain using equation \ref{eqInfGain}:
\begin{equation*}
\begin{split}
IG(D_{\text{Server}}) &= 0.971 - \frac{2}{10} \times 0. + \frac{2}{10} \times 0. \\
&\quad + \frac{1}{10} \times 0. + \frac{5}{10} \times 0.971 \\
&= 0.4855
\end{split}
\end{equation*}
by applying the same step to the Geographic Location attribute, we can get an information gain of:
\begin{equation*}
\begin{split}
IG(D_{\text{GL}}) &= 0.971 - \frac{2}{10} \times 0. + \frac{3}{10} \times 0 . 917666 \\
&\quad + \frac{3}{10} \times 0 . 917666 + \frac{2}{10} \times 0. \\
&= 0.4204004
\end{split}
\end{equation*}
By following this approach, we can compute the entropy values for the attributes and further analyze the newcomer federated servers.

\subsubsection{Trust Score Aggregartion}

After setting up the recommendation trees on each side of the recommender system and obtaining the recommendations, we can move forward to the next phase, which is endorsement aggregation. The main objective of this phase is to combine the collected endorsements in a way that can handle dishonest and misleading submissions. To accomplish this, we apply the Dempster-Shafer method of evidence, which is known for its ability to combine information from different sources. What makes our approach different from existing techniques for trust establishment is that it doesn't rely on predetermined thresholds for decision-making. Instead, we calculate beliefs about both untrustworthiness and trustworthiness of devices and compare them to make the final decision. Additionally, our approach assigns weights to endorsers based on their credibility scores, which are constantly updated to reflect changes in their honesty over time. This dynamic weight assignment approach differs from the static weight assignment used in other methods.

Let \(\mathcal{H} = \{H_1, H_2, H_3\}\) represent the set of three hypotheses that correspond to different endorsements for a newly deployed server. In this context, \(H_1\) denotes trustworthiness, \(H_2\) denotes untrustworthiness, and \(H_3\) signifies uncertainty between distrust and trust. The basic probability assignment ($BPA$) \(m_{i}^{b}(H)\) for a specific hypothesis \(H\) about a server \(i\) given by bootstrapper \(b\) correlates with the credibility score of \(b\).

In particular, when bootstrapper \(b\) attributes a trustworthiness score of \(\lambda\) to server \(i\), the calculation of the $BPA$s for various hypotheses proceeds as follows:

\begin{itemize}
    \item \(m_{i}^{b}(H_1) = \lambda\)
    \item \(m_{i}^{b}(H_2) = 0\)
    \item \(m_{i}^{b}(H_3) = 1 - \lambda\)
\end{itemize}

Alternatively, if bootstrapper \(b\) determines server \(i\) as untrustworthy, the $BPA$s are calculated as:

\begin{itemize}
    \item \(m_{i}^{b}(H_1) = 0\)
    \item  \(m_{i}^{b}(H_2) = \lambda\)
    \item \(m_{i}^{b}(H_3) = 1 - \lambda\)
\end{itemize}

Once all the $BPA$s are defined, the ultimate collective belief function concerning a specific hypothesis \(H\) is derived by adding together the $BPA$s contributed by various endorsing bootstrappers for \(H\).
The belief function determined by recommender system \(r\) concerning the trustworthiness of server \(i\) after consulting two bootstrappers \(b'\) and \(b\) is expressed as:

\begin{equation}
\begin{aligned}
\theta_{i}^{r}(H_1) &= \frac{1}{K} \biggl[m_{i}^{b}(H_1) \cdot m_{i}^{b'}(H_1) \\
&\quad+ m_{i}^{b}(H_1) \cdot m_{i}^{b'}(H_3) + m_{i}^{b}(H_3) \cdot m_{i}^{b'}(H_1)\biggr]
\end{aligned}
\end{equation}

Similarly, after consulting two bootstrappers \(b'\) and \(b\) regarding the untrustworthiness of item \(i\)'s the belief function computed by \(r\) is provided by:

\begin{equation}
\begin{aligned}
\theta_{i}^{r}(H_2) &= \frac{1}{K} \biggl[m_{i}^{b}(H_2) \cdot m_{i}^{b'}(H_2) \\
&\quad+ m_{i}^{b}(H_2) \cdot m_{i}^{b'}(H_3) + m_{i}^{b}(H_3) \cdot m_{i}^{b'}(H_2)\biggr]
\end{aligned}
\end{equation}

Finally, after consulting bootstrappers \(b'\) and \(b\), the belief function determined by \(r\) for the federated server \(i\) with respect to its trustworthiness or untrustworthiness (i.e., overall uncertainty) is represented as:

\begin{equation}
\theta_{i}^{r}(H_3) = \frac{1}{K} \left[m_{i}^{b}(H_3) \cdot m_{i}^{b'}(H_3)\right]
\
\end{equation}

Where:
\begin{equation}
K = \sum_{h \cap h' = \emptyset} m_i^b(h) \cdot m_i^{b'}(h')
\end{equation}

The values generated by various belief functions are real numerical values falling within the range of 0 to 1, meaning $\theta_i^r(H1)$, $\theta_i^r(H2)$, and $\theta_i^r(H3)$ all lie within the interval [0, 1]. In conclusion, the recommender system's decision regarding a given newcomer server $i$ is determined by evaluating the beliefs in its lack of trustworthiness $\theta_i^r(H3)$ and trustworthiness $\theta_i^r(H1)$ and subsequently comparing these values. In other words, if $\theta_i^r(H1) > \theta_i^r(H3)$, server $i$ is classified as trustworthy; otherwise, server $i$ is considered untrustworthy.

\subsection{Credibility Score}

The credibility score adjustment procedure, described by Equation (\ref{crup}), guarantees that each bootstrapper's credibility score is updated relatively to the contrast between their computed endorsement $R(u, i)$ for the federated server $i$ and the decision produced by the DST-based aggregation process. This update process rewards users whose endorsements align with the final decision, increasing their credibility scores, while users whose endorsements deviate from the final decision experience a decrease in their credibility scores. Maintaining accurate credibility scores is crucial for the bootstrapping process honesty, as the effectiveness of the DST aggregation method heavily relies on the endorsers' credibility scores.

\begin{equation}
\label{crup}
\phi(r \rightarrow u) = 
\begin{cases}
\min(1, \phi(r \rightarrow u) + X), & \text{if } C1 \\
|\phi(r \rightarrow u) - Y|, & \text{if } C2
\end{cases}
\end{equation}

where 
\begin{itemize}
    \item $X = \max(\theta_{ir}(T),\theta_{ir}(N))$
    \item $Y = \min(\theta_{ir}(T), \theta_{ir}(N))$
\end{itemize}

Accordingly, the conditions $C1$ and $C2$ are defined as follows:

\begin{itemize}
    \item \textbf{C1:} $R(u, i) \in \{T\} \land \theta_{ir}(T) > \theta_{ir}(N)$ or $R(u, i) \in \{N\} \land \theta_{ir}(T) < \theta_{ir}(M)$
    \item \textbf{C2:} $R(u, i) \in \{T\} \land \theta_{ir}(T) < \theta_{ir}(N)$ or $R(u, i) \in \{N\} \land \theta_{ir}(T) > \theta_{ir}(N)$
\end{itemize}

\section{Client Selection: A Mutual Approach}
\label{proposed}

In this section, we describe the introduced framework, explain the federated server and the client preference functions, and provide the mutual selection algorithms for the client and server.

\subsection{Mutual Trust Matching approach Overview}
\label{}

\begin{figure}[ht]
    \centering
    \includegraphics[width=0.9 \linewidth]{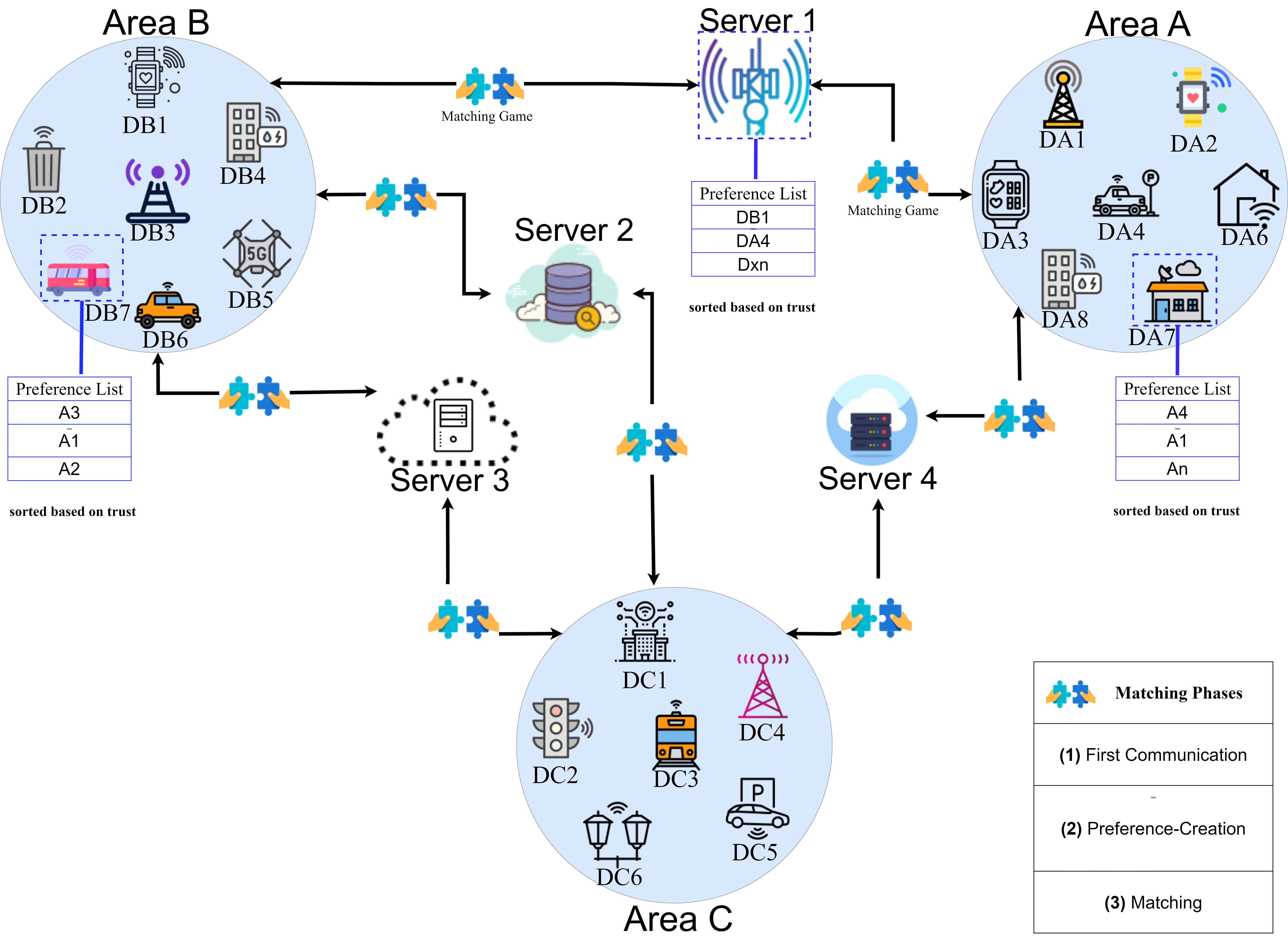}
    \caption{Overview of the Proposed Method}
    \label{Archi}
\end{figure}

Our proposed approach requires two distinct sets of inputs: a group of operational smart devices labelled as $D$ that are ready and motivated to participate in the Federated Learning (FL) model training process and a group of federated servers denoted as $A$ that must select a group of clients $D\prime \subset D$ for executing federated learning tasks. In Figure \ref{Archi}, you can observe the high-level architecture of our strategy within a smart urban environment.

The proposed approach comprises three primary stages. (1) Initially, federated servers communicate with smart devices regarding their operational requirements. (2) In the phase of preference establishment, each federated server and client construct their preference array based on their trust score. (3) Subsequently, federated servers and clients are matched together.

\subsection{Mutual Trust Game Theory Formulation}
\label{matchingametheoryformulation}

This section outlines the mutual trust game and the process of generating preference lists for federated servers and clients.

\textbf{Definition 1}: Consider the matching relation $\Gamma$ between IoT/IoV devices and federated servers established through the matching game. $\Gamma$ is represented as a function $A \cup D \rightarrow 2^{A \cup D}$, and it Comply to the following conditions:
\begin{itemize}
    \item For each client $d$, $\Gamma(d)$ is a subset of $A$. Where $|\Gamma(d)| = 0$ indicates that client $d$ remains unallocated to any federated server.
    \item For each federated server $a$, $\Gamma(a)$ is a subset of $D$. In cases where $S_{a} < K_a$, it signifies that the required number of smart devices for federated server $a$ in its FL task has not been met.
    \item The allocation relationship between client $d$ and federated server $a$, i.e., $d \in \Gamma(a)$, holds true if and only if $\Gamma(d) = a$ for all $d \in D$ and $a \in A$.
\end{itemize}

\textbf{Definition 2}: A relation between Server-Client $\Gamma$ is termed "restricted" if there exists a pair ($a, d$) where $a$ belongs to $\Gamma(d)$ and $d$ belongs to $\Gamma(a)$. Additionally, the conditions $a \succ_d \Gamma(d)$ and $d \succ_a \Gamma(a)$ must be satisfied.

\textbf{Definition 3}: A federated server, denoted as $a$, is considered satisfied when it reaches the desired number of clients, denoted as $K_a$. If a server still needs clients, it will accept any smart device that meets the condition $S_a < k_a$.

\textbf{Definition 4}: The $\Gamma$ relation is considered stable under the following conditions: (1) The relationship is not restricted, and (2) each federated server has a sufficient number of client devices in $S_{a}$ to meet its requirements.

\subsection{Preference Function: Federated Servers}
\label{ServerMatching}

Federated servers mainly aim to enhance the training accuracy of deep learning models by carefully selecting client devices with the highest trust scores. Each federated server $a \in A$ has a transitive, strict, and complete preference relation $P_a(D)$ with each smart device $d \in D$. If a preference relationship $d^\prime \succ_a d^{\prime\prime}$ exists, federated server $a$ prefers smart device $d^\prime$ over smart device $d^{\prime\prime}$. Furthermore, if $a$ does not obviously prefer $d$ to stay unpaired, the smart device $d$ is categorized as unappealing to $a$.

\subsection{Preference Function: Clients}
\label{IoTMatching}

Clients aim to be matched to the federated servers with the highest trust score. With each smart device $d \in D$, each federated server $a \in A$ has a transitive, strict, and complete preference connection $P_d(A)$. If a preference relation $a \succ_d a^\prime$ exists, federated server $a$ is preferred over federated server $a^\prime$ for a smart device $d$. Furthermore, a federated server $a$ is labelled undesirable to $d$ if a smart device $d$ does not clearly prefer joining $a$ or staying unpaired.

\subsection{Matching Algorithms: Clients \& Federated Server }
\label{matchingselection}

The following phase entails creating the appropriate algorithms to carry out the intended matching based on the generated preference lists obtained from Algorithm ~\ref{client-pref} and the bootstrapping method introduced in (section~\ref{smartrust}). Since our methodology is distributed, we introduce two separate algorithms (\ref{iot_matching_alg} and \ref{fog_matching_alg}). Algorithm~\ref{iot_matching_alg} operates on each smart device, while Algorithm~\ref{fog_matching_alg} operates on each federated server. This procedure results in the association of clients with federated servers. To elaborate on Algorithm~\ref{iot_matching_alg}, each client device starts a pairing process with its designated federated server, guided by its preference list. On the contrary, through Algorithm~\ref{fog_matching_alg}, each federated server responds to smart devices with acceptance or rejection decisions according to its preference list.

\begin{algorithm}[ht]
\caption{Selection Process - Smart Devices}
\label{iot_matching_alg}
\begin{algorithmic}[1]
\Require {Array $P_d$ containing client preferences}
\Ensure {Matching relation between client device and federated server}
\Repeat
\ForEach {$a \in P_d$}
\State{Send pairing request $message_{d \rightarrow m}$ to $a$}
\State{Await for the response $Response_{m \rightarrow d}$ from $a$}
\State Label $a$ as visited
\If{$Response_{a \rightarrow d}$ == "Accept"}
    \State{Match $d$ and $a$}
\EndIf
\EndFor
\State{\textbf{Rebuild} the preference array of smart device }
\Until{Reach the end of the federated training process}
\end{algorithmic}
\end{algorithm}

Algorithm \ref{iot_matching_alg} uses the federated server preference list generated based on the method discussed in Section~\ref{smartrust}. It then iterates over each smart device preference list (Line $2$), determining the most preferred server. The smart device then initiates a pairing request to the designated server (Line $3$), anticipating a response (Line $4$), then marking the server as visited (Line $5$). If the server approves, the server and smart device are matched (Lines $6-9$). If the smart device receives a negative response, it will relegate the server to the end of the preference list. This technique is repeated for each round of federated learning (Line $11$). The method discussed in Section~\ref{smartrust} is run within Algorithm \ref{iot_matching_alg} (Line $10$) to acquire updated preference lists that consider smart devices' performance changes and the additions or removals of servers.
Algorithm \ref{iot_matching_alg} exhibits a linear relationship with every round, denoted by $O(n)$, where $n$ represents the count of federated servers within the preference list $P_i$, since it evaluates each server's availability for federated learning.

\begin{algorithm}[ht]
\caption{Selection Process - Federated Server}
\label{fog_matching_alg}
\begin{algorithmic}[1]
\Require {A list $R_d$ of client devices with pairing requests to server $a$}
\Ensure {Smart devices matched to a specific server $a$}
\Repeat
\While{$R_i$ not empty}
\If{$S_a < K_a$}
    \State{Send acceptance response $\rightarrow Reply_{a \rightarrow d}$ ='Accept'}
    \State{Include $d$ in the list $L_{S_a}$}
    \State{increment $S_a$ by 1}
\ElsIf{$d \succ_a d^\prime $ \textbf{and} $S_a == K_a$}
    \State{Send rejection response $Respond_{a \rightarrow d^\prime}$ = 'Reject'}
    \State{Remove $d^\prime$ from the list $L_{S_a}$}
    \State{Send accept response $\rightarrow Respond_{s \rightarrow d}$ = 'Accept'}
    \State{Include $d$ in the list $L_{S_a}$}
\Else
    \State{Send rejection response to $d \rightarrow Respond_{a \rightarrow d}$ = 'Reject'}
    \State{Decline every $d^\prime \in R_d$ where $d \succ_a d^\prime $}
\EndIf
\EndWhile
\State{\textbf{Rebuild} the federated server preference array}
\Until{Reach the end of the federated training process}
\end{algorithmic}
\end{algorithm}

Each federated server implements Algorithm \ref{fog_matching_alg} to manage an input queue of smart devices. This algorithm verifies the queue's emptiness (Line $2$). When there are outstanding requests, It verifies whether the target number of clients has been achieved (Line $3$). If not, the server issues an acknowledgment to the smart device (Line $4$), increases the count of selected clients, and includes the client in the list (Lines $5-6$). Nevertheless, in the event that the desired number of clients is achieved, and the current smart device has a higher rating than any previously chosen device (Line $7$), The server ends the contract with the device of low rank (Lines $8-9$). Subsequently, the server dispatches an \textit{Accept} response to the presently chosen device, adding it to the list (Lines $10-11$). If the current smart device does not have a better rating than any already chosen device, a \textit{reject} response is transmitted (Line $13$). In this scenario, the server removes lower-ranked devices (Line $14$). This procedure recurs in each training round (Line $18$), with Algorithm \ref{client-pref} performing preference list updates in response to environmental changes (Line $17$). For a single round, Algorithm \ref{fog_matching_alg} exhibits a time complexity of $O(n \log n)$, where $n$ corresponds to the number of smart devices in the queue $Q_i$. To select suitable devices for the server, The algorithm traverses every device in the queue, efficiently comparing device priorities using either a balanced search tree or a priority queue. The time spent receiving and sending messages and awaiting responses remains modest and consistent.

\section{Simulation \& Experimental Results}
\label{experiments}

In this section, we describe the environment in which our simulations were conducted and analyze the outcomes of our experiments.

\subsection{Experiment Setup}
\label{experimentalsetup}

The initial environment comprises a set of 200 client devices, each associated with dataset sizes varying from [100, 250] images. These smart devices are assigned between three and four labels from a total of ten available labels. In order to carry out our simulations, we utilized two datasets: the MNIST dataset from the National Institute of Standards and Technology (NIST) \footnote{http://yann.lecun.com/exdb/mnist/} and the Fashion MNIST dataset\footnote{\text{https://www.tensorflow.org/datasets/catalog/fashion\_mnist}}. These datasets were thoughtfully distributed across the smart devices using a non-IID approach. To conduct the experiments, we built upon the ModularFed environment~\cite {arafeh} as our foundation. Each of the smart devices has varying resource capabilities, with RAM capacity ranging from [100, 1300] megabytes, CPU capacity ranging from [50,1000] million instructions per second (MIPS), and network bandwidth varying from [50, 1300] megabits per second (Mbps). For our experimentation, we established two federated servers, each requesting K=20 clients, and intended to conduct a total of 100 federated learning (FL) training rounds, denoted as R. To evaluate the efficacy of our approach. We compared it to the original federated learning technique, which was initially introduced by Google~\cite{p1}. In our experiments, we refer to this conventional method as "VanillaFL."

\subsection{Simulation Outcomes}
\label{experimentalresults}
Our simulations are primarily intended to explore two critical metrics: (I) the selection of untrustworthy clients and (II) the server's global model's overall accuracy. Throughout this section, Servers (1 and 2) will symbolize the servers operating within our technique, while Server1R and Server2R will act as the servers applying the VanillaFL method.

\begin{figure}[ht]
    \centering
    \includegraphics[width=\linewidth]{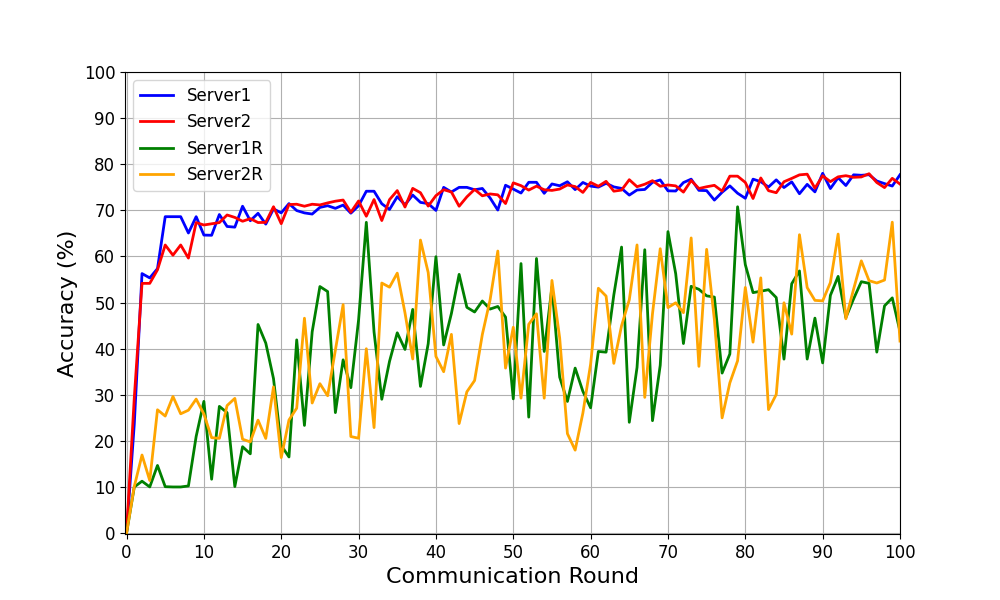}
    \caption{Accuracy of Global Model vs. Number of Rounds: Fashion MNIST Dataset.}
    \label{fig:r1}
\end{figure}

We used the Fashion MNIST dataset in Figure~\ref{fig:r1} to investigate the global model accuracy of our technique to the VanillaFL in the existence of untrustworthy clients in the environment. Every round, we incorporate an additional 10 smart devices into the initial set of smart devices in the environment. The graph shows that our technique, shown by the blue and red lines, surpasses the VanillaFL approach by more than 15\% in the first round on both servers. Servers 1 and 2 reached an accuracy of 84.71\% and 84.69\%, respectively. In the 100th round, our approach exceeded VanillaFL by more than 38\%. Furthermore, the variation in the Vanilla server's lines emphasizes the vulnerability of the random technique.

\begin{figure}[ht]
    \centering
    \includegraphics[width=\linewidth]{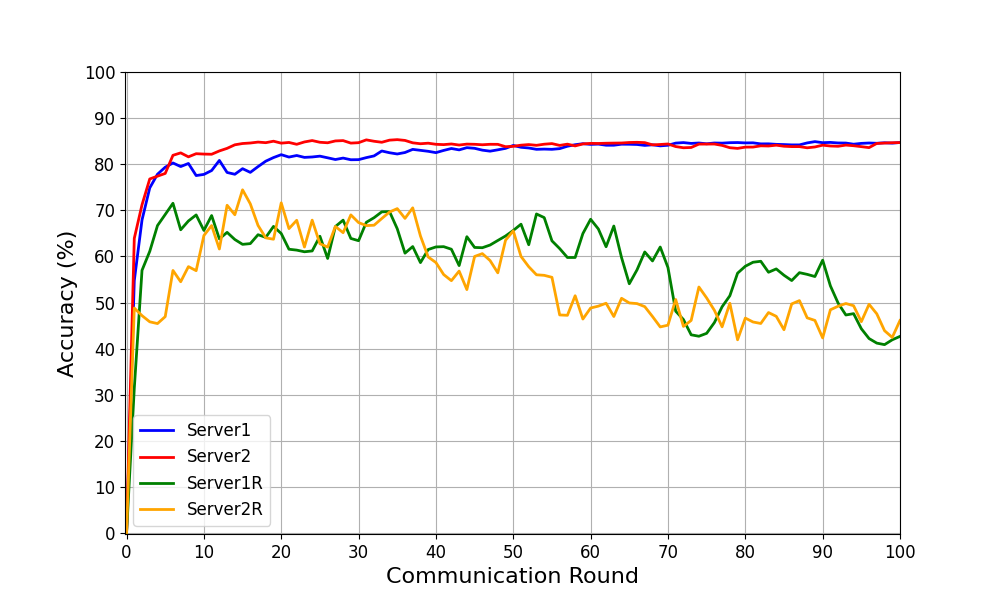}
    \caption{Accuracy of Global Model vs. Number of Rounds: MNIST Dataset.}
    \label{fig:r2}
\end{figure}

In Figure~\ref{fig:r2}, we compared our method and VanillaFL using the MNIST dataset. The evaluation focused on global model accuracy amidst the presence of non-trustworthy devices. Our approach consistently outperformed VanillaFL, displaying over 13\% higher accuracy from the first round. By the 100th round, servers achieved remarkable accuracy of 77.71\% and 75.69\%, respectively, surpassing VanillaFL by more than 33\%. These findings underscore the effectiveness of our method in challenging scenarios with diverse participants.

\begin{figure}[ht]
    \centering
    \includegraphics[width=\linewidth]{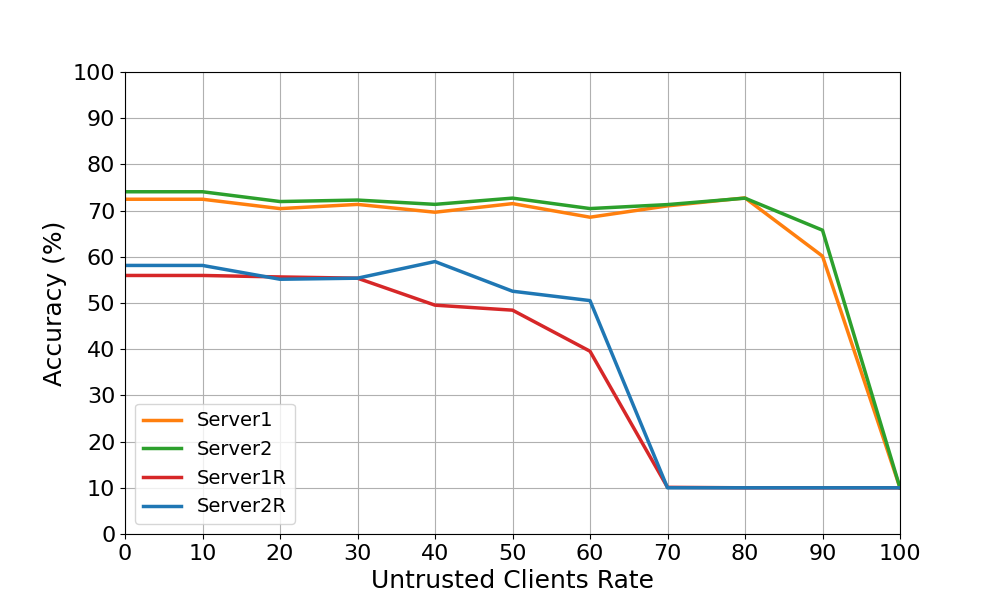}
    \caption{Accuracy Vs. Untrusted Client Rate: MNIST Dataset.}
    \label{fig:r3}
\end{figure}

In Figure~\ref{fig:r3}, we analyze the effectiveness of our method towards the VanillaFL while incrementing the proportion of untrustworthy clients in the given setting using the Mnist dataset. As depicted in the graph, our approach, illustrated by servers 1 and 2, surpasses the performance of the VanillaFL method, even in environments with a significant number of untrustworthy smart devices. Our technique consistently maintains an accuracy of approximately 72\%, even when 80\% of the devices are untrustworthy, meaning that only 40 smart devices are considered trustworthy. In contrast, the accuracy of VanillaFL experiences a rapid decline as the portion of untrustworthy clients increases, falling to nearly 10\% when 80\% of the clients are untrustworthy.

\begin{figure}[ht]
    \centering
    \includegraphics[width=\linewidth]{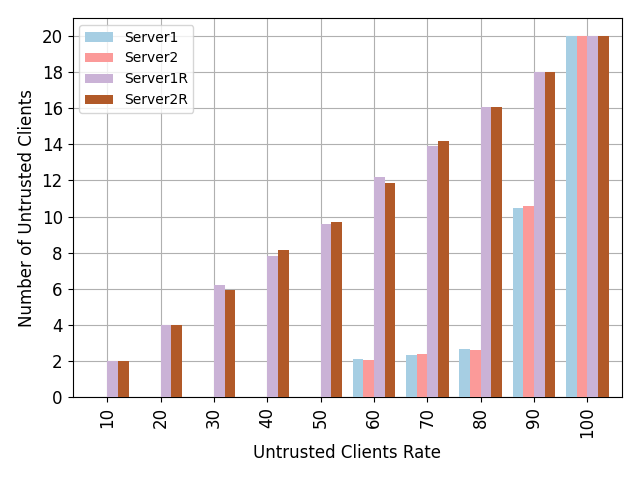}
    \caption{Untrusted Client Selected Vs. Untrusted Client Rate: MNIST Dataset.}
    \label{fig:r4}
\end{figure}

Figure~\ref{fig:r4}, shows a comparison of the average selection of untrustworthy clients in 100 rounds as we progressively increase the portion of untrustworthy devices in the framework. The bar plot illustrates that our strategy leads to a substantial reduction in the average count of untrustworthy selected clients, even when confronted with a highly untrustworthy environment consisting of 80\% untrustworthy devices, leaving only 40 clients in the trusted category. Our approach results in an average selection of approximately 2.68 untrustworthy smart devices out of 20 per server, a significant improvement compared to the VanillaFL approach, which averages 16.06 untrustworthy devices out of 20 per server.

\begin{figure}[ht]
    \centering
    \includegraphics[width=\linewidth]{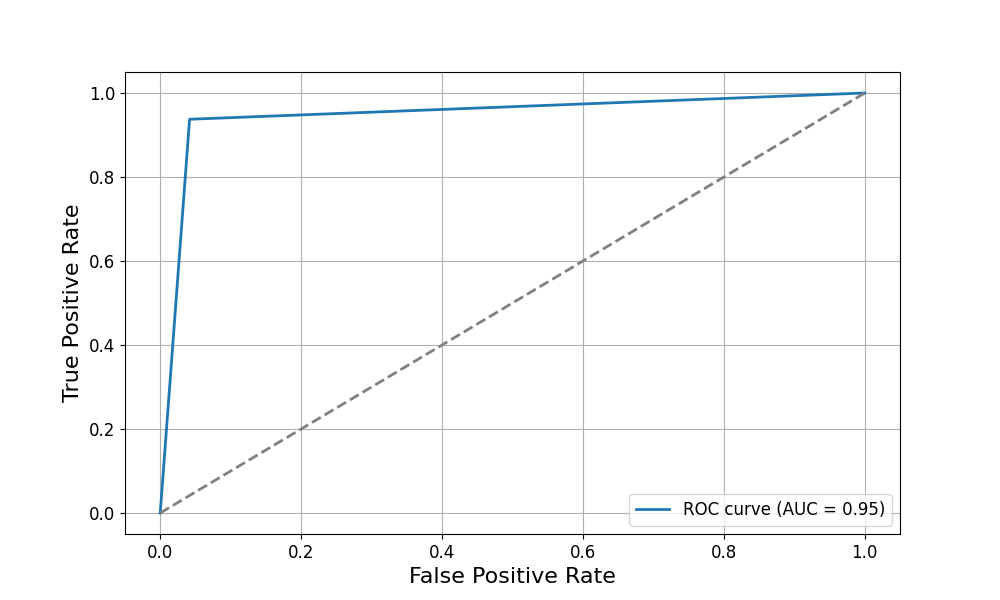}
    \caption{Bootstrapping Efficiency Based on ROC Curve and AUC.}
    \label{fig:r5}
\end{figure}

Figure~\ref{fig:r5} illustrates the receiver operating characteristic (ROC) curve data, presenting the trade-off between the false positive and the true positive rate. The area under the curve (AUC) is 0.95, showing that the classes are perfectly separated. The ROC curve, particularly, has a high true positive rate of 0.9375 and a relatively low false positive rate of 0.0417. These findings highlight the classifier's ability in binary classification tasks, making it a suitable option for applications requiring high discriminating performance.

\section{Conclusion}
\label{Conclusion}

In this research, we presented innovative matching-based game theory-based strategies for addressing the client-server selection dilemma in federated learning settings, highlighting the necessity of mutual trust and authentication for IoT/IoV devices and federated servers. Our technique improves the overall accuracy of the global federated learning model and reduces the selection of untrustworthy clients, resulting in fewer perceived interactions.
The simulation outcomes prove the effectiveness of our mutual trust matching-based selection technique, resulting in a significant boost in global model accuracy of more than 33\% as compared to the conventional VanillaFL on two varied datasets. Furthermore, in an 80\% malevolent environment, our strategy decreases the average number of untrustworthy clients chosen from 16 to 2 out of 20.
In the future, we will focus on a full investigation of the network's sensitivity to device collusion to mitigate or prevent it entirely. Furthermore, we are working to improve a self-healing system that automatically increases the number of trusted clients inside the network while requiring less time and data. These developments can potentially improve the security and efficacy of federated learning systems.

\bibliographystyle{IEEEtran}
\bibliography{references}

\begin{thebibliography}{10}
\providecommand{\url}[1]{#1}
\csname url@samestyle\endcsname
\providecommand{\newblock}{\relax}
\providecommand{\bibinfo}[2]{#2}
\providecommand{\BIBentrySTDinterwordspacing}{\spaceskip=0pt\relax}
\providecommand{\BIBentryALTinterwordstretchfactor}{4}
\providecommand{\BIBentryALTinterwordspacing}{\spaceskip=\fontdimen2\font plus
\BIBentryALTinterwordstretchfactor\fontdimen3\font minus \fontdimen4\font\relax}
\providecommand{\BIBforeignlanguage}[2]{{%
\expandafter\ifx\csname l@#1\endcsname\relax
\typeout{** WARNING: IEEEtran.bst: No hyphenation pattern has been}%
\typeout{** loaded for the language `#1'. Using the pattern for}%
\typeout{** the default language instead.}%
\else
\language=\csname l@#1\endcsname
\fi
#2}}
\providecommand{\BIBdecl}{\relax}
\BIBdecl

\bibitem{hani}
H.~Sami, R.~Saado, A.~El~Saoudi, A.~Mourad, H.~Otrok, and J.~Bentahar, ``Opportunistic uav deployment for intelligent on-demand iov service management,'' \emph{IEEE Transactions on Network and Service Management}, 2023.

\bibitem{mario}
M.~Chahoud, S.~Otoum, and A.~Mourad, ``On the feasibility of federated learning towards on-demand client deployment at the edge,'' \emph{Information Processing \& Management}, vol.~60, no.~1, p. 103150, 2023.

\bibitem{rabab1}
H.~Abualola, R.~Mizouni, H.~Otrok, S.~Singh, and H.~Barada, ``A matching game-based crowdsourcing framework for last-mile delivery: Ground-vehicles and unmanned-aerial vehicles,'' \emph{Journal of Network and Computer Applications}, p. 103601, 2023.

\bibitem{adoom2}
X.~Wang, S.~Garg, H.~Lin, J.~Hu, G.~Kaddoum, M.~J. Piran, and M.~S. Hossain, ``Toward accurate anomaly detection in industrial internet of things using hierarchical federated learning,'' \emph{IEEE Internet of Things Journal}, vol.~9, no.~10, pp. 7110--7119, 2021.

\bibitem{sarhad}
S.~Arisdakessian, O.~A. Wahab, A.~Mourad, and H.~Otrok, ``Coalitional federated learning: Improving communication and training on non-iid data with selfish clients,'' \emph{IEEE Transactions on Services Computing}, 2023.

\bibitem{adoom1}
G.~Rathee, S.~Garg, G.~Kaddoum, B.~J. Choi, M.~Hassan, and S.~A. Alqahtani, ``Trustsys: Trusted decision making scheme for collaborative artificial intelligence of things,'' \emph{IEEE Transactions on Industrial Informatics}, 2022.

\bibitem{waz}
M.~Wazzeh, H.~Ould-Slimane, C.~Talhi, A.~Mourad, and M.~Guizani, ``Warmup and transfer knowledge-based federated learning approach for iot continuous authentication,'' \emph{arXiv preprint arXiv:2211.05662}, 2022.

\bibitem{jamal}
A.~Alagha, S.~Singh, R.~Mizouni, J.~Bentahar, and H.~Otrok, ``Target localization using multi-agent deep reinforcement learning with proximal policy optimization,'' \emph{Future Generation Computer Systems}, vol. 136, pp. 342--357, 2022.

\bibitem{i1}
S.~AbdulRahman, H.~Tout, H.~Ould-Slimane, A.~Mourad, C.~Talhi, and M.~Guizani, ``A survey on federated learning: The journey from centralized to distributed on-site learning and beyond,'' \emph{IEEE Internet of Things Journal}, vol.~8, no.~7, pp. 5476--5497, 2020.

\bibitem{ahmad}
A.~Hammoud, A.~Mourad, H.~Otrok, and Z.~Dziong, ``Data-driven federated autonomous driving,'' in \emph{Mobile Web and Intelligent Information Systems: 18th International Conference, MobiWIS 2022, Rome, Italy, August 22--24, 2022, Proceedings}.\hskip 1em plus 0.5em minus 0.4em\relax Springer, 2022, pp. 79--90.

\bibitem{mohsen}
A.~Tabassum, A.~Erbad, W.~Lebda, A.~Mohamed, and M.~Guizani, ``Fedgan-ids: Privacy-preserving ids using gan and federated learning,'' \emph{Computer Communications}, vol. 192, pp. 299--310, 2022.

\bibitem{i3}
P.~Kairouz, H.~B. McMahan, B.~Avent, A.~Bellet, M.~Bennis, A.~N. Bhagoji, K.~Bonawitz, Z.~Charles, G.~Cormode, R.~Cummings \emph{et~al.}, ``Advances and open problems in federated learning,'' \emph{Foundations and Trends{\textregistered} in Machine Learning}, vol.~14, no. 1--2, pp. 1--210, 2021.

\bibitem{guesmi2023physical}
A.~Guesmi, M.~A. Hanif, B.~Ouni, and M.~Shafique, ``Physical adversarial attacks for camera-based smart systems: Current trends, categorization, applications, research challenges, and future outlook,'' \emph{IEEE Access}, 2023.

\bibitem{omar}
O.~A. Wahab, ``Intrusion detection in the iot under data and concept drifts: Online deep learning approach,'' \emph{IEEE Internet of Things Journal}, vol.~9, no.~20, pp. 19\,706--19\,716, 2022.

\bibitem{p1}
B.~McMahan, E.~Moore, D.~Ramage, S.~Hampson, and B.~A. y~Arcas, ``Communication-efficient learning of deep networks from decentralized data,'' in \emph{Artificial intelligence and statistics}.\hskip 1em plus 0.5em minus 0.4em\relax PMLR, 2017, pp. 1273--1282.

\bibitem{r6}
O.~Wehbi, S.~Arisdakessian, O.~A. Wahab, H.~Otrok, S.~Otoum, A.~Mourad, and M.~Guizani, ``Fedmint: Intelligent bilateral client selection in federated learning with newcomer iot devices,'' \emph{arXiv preprint arXiv:2211.01805}, 2022.

\bibitem{r7}
O.~Wehbi, S.~Arisdakessian, O.~A. Wahab, H.~Otrok, S.~Otoum, and A.~Mourad, ``Towards bilateral client selection in federated learning using matching game theory,'' in \emph{GLOBECOM 2022-2022 IEEE Global Communications Conference}.\hskip 1em plus 0.5em minus 0.4em\relax IEEE, 2022, pp. 01--06.

\bibitem{r1}
G.~Rjoub, O.~Abdel~Wahab, J.~Bentahar, and A.~Bataineh, ``A trust and energy-aware double deep reinforcement learning scheduling strategy for federated learning on iot devices,'' in \emph{Service-Oriented Computing: 18th International Conference, ICSOC 2020, Dubai, United Arab Emirates, December 14--17, 2020, Proceedings 18}.\hskip 1em plus 0.5em minus 0.4em\relax Springer, 2020, pp. 319--333.

\bibitem{r2}
N.~Chen, Y.~Li, X.~Liu, and Z.~Zhang, ``A mutual information based federated learning framework for edge computing networks,'' \emph{Computer Communications}, vol. 176, pp. 23--30, 2021.

\bibitem{r3}
X.~Cao, M.~Fang, J.~Liu, and N.~Z. Gong, ``Fltrust: Byzantine-robust federated learning via trust bootstrapping,'' \emph{arXiv preprint arXiv:2012.13995}, 2020.

\bibitem{r4}
O.~A. Wahab, G.~Rjoub, J.~Bentahar, and R.~Cohen, ``Federated against the cold: A trust-based federated learning approach to counter the cold start problem in recommendation systems,'' \emph{Information Sciences}, vol. 601, pp. 189--206, 2022.

\bibitem{r5}
X.~Bao, C.~Su, Y.~Xiong, W.~Huang, and Y.~Hu, ``Flchain: A blockchain for auditable federated learning with trust and incentive,'' in \emph{2019 5th International Conference on Big Data Computing and Communications (BIGCOM)}.\hskip 1em plus 0.5em minus 0.4em\relax IEEE, 2019, pp. 151--159.

\bibitem{r8}
X.~Cao, M.~Fang, J.~Liu, and N.~Z. Gong, ``Fltrust: Byzantine-robust federated learning via trust bootstrapping,'' \emph{arXiv preprint arXiv:2012.13995}, 2020.

\bibitem{nishio2019client}
T.~Nishio and R.~Yonetani, ``Client selection for federated learning with heterogeneous resources in mobile edge,'' in \emph{ICC 2019-2019 IEEE international conference on communications (ICC)}.\hskip 1em plus 0.5em minus 0.4em\relax IEEE, 2019, pp. 1--7.

\bibitem{abdulrahman2020fedmccs}
S.~AbdulRahman, H.~Tout, A.~Mourad, and C.~Talhi, ``Fedmccs: Multicriteria client selection model for optimal iot federated learning,'' \emph{IEEE Internet of Things Journal}, vol.~8, no.~6, pp. 4723--4735, 2020.

\bibitem{r9}
Y.~Dong, X.~Chen, K.~Li, D.~Wang, and S.~Zeng, ``Flod: Oblivious defender for private byzantine-robust federated learning with dishonest-majority,'' in \emph{Computer Security--ESORICS 2021: 26th European Symposium on Research in Computer Security, Darmstadt, Germany, October 4--8, 2021, Proceedings, Part I}, 2021, pp. 497--518.

\bibitem{r11}
L.~Zhang, T.~Zhu, P.~Xiong, W.~Zhou, and S.~Y. Philip, ``A robust game-theoretical federated learning framework with joint differential privacy,'' \emph{IEEE Transactions on Knowledge and Data Engineering}, vol.~35, no.~4, pp. 3333--3346, 2022.

\bibitem{r12}
D.~Chen, C.~S. Hong, L.~Wang, Y.~Zha, Y.~Zhang, X.~Liu, and Z.~Han, ``Matching-theory-based low-latency scheme for multitask federated learning in mec networks,'' \emph{IEEE Internet of Things Journal}, vol.~8, no.~14, pp. 11\,415--11\,426, 2021.

\bibitem{r13}
J.~Kang, Z.~Xiong, D.~Niyato, Z.~Cao, and A.~Leshem, ``Training task allocation in federated edge learning: A matching-theoretic approach,'' in \emph{2020 IEEE 17th Annual Consumer Communications \& Networking Conference (CCNC)}.\hskip 1em plus 0.5em minus 0.4em\relax IEEE, 2020, pp. 1--6.

\bibitem{pf1}
O.~A. Wahab, R.~Cohen, J.~Bentahar, H.~Otrok, A.~Mourad, and G.~Rjoub, ``An endorsement-based trust bootstrapping approach for newcomer cloud services,'' \emph{Information Sciences}, vol. 527, pp. 159--175, 2020.

\bibitem{wahab2016towards}
O.~A. Wahab, J.~Bentahar, H.~Otrok, and A.~Mourad, ``Towards trustworthy multi-cloud services communities: A trust-based hedonic coalitional game,'' \emph{IEEE Transactions on Services Computing}, vol.~11, no.~1, pp. 184--201, 2016.

\bibitem{quinlan1986induction}
J.~R. Quinlan, ``Induction of decision trees,'' \emph{Machine learning}, vol.~1, pp. 81--106, 1986.

\bibitem{arafeh}
M.~Arafeh, H.~Otrok, H.~Ould-Slimane, A.~Mourad, C.~Talhi, and E.~Damiani, ``Modularfed: Leveraging modularity in federated learning frameworks,'' \emph{Internet of Things}, p. 100694, 2023.

\end{thebibliography}

\begin{IEEEbiography}[{\includegraphics[width=1in,height=1.25in,clip,keepaspectratio]{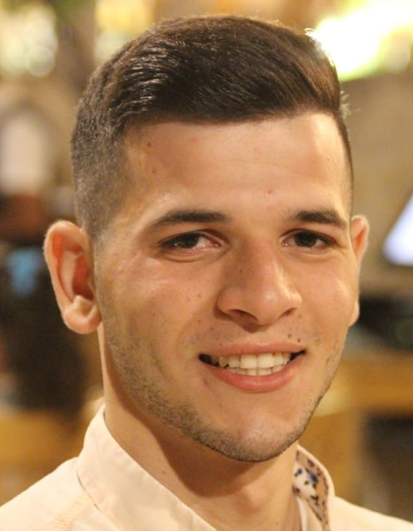}}]%
{\textbf{\textit{\textbf{Osama Wehbi}}}}
is currently working as a Research Assistant at the Cyber Security Systems and Applied AI Research Center, Department of CSM, Lebanese American University, Lebanon, and Mohammad Bin Zayed University of Artificial Intelligence, Abu Dhabi, UAE. He holds a Master's degree in Business Computing from the Lebanese University and an M.Sc in Computer Science from the Lebanese American University. With previous experience as a Research and Teaching Assistant, his research interests encompass areas such as Federated Machine Learning, Game Theory, Federated Cloud and Fog, Artificial Intelligence, and Cyber Security. 
\end{IEEEbiography}

\vskip 0pt plus -1fil

\begin{IEEEbiography}[{\includegraphics[width=1in,height=1.25in,clip,keepaspectratio]{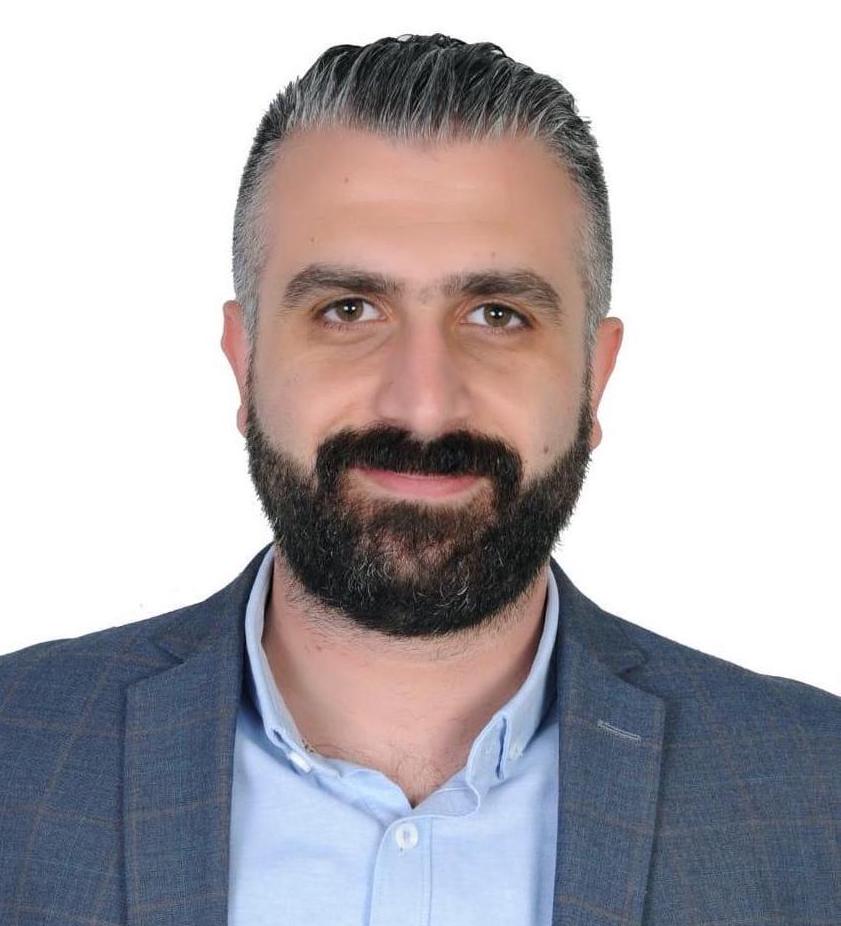}}]%
{\textbf{\textit{\textbf{Sarhad Arisdakessian}}}}
is a Ph.D. student in Computer Science in the Department of Computer Engineering at Polytechnique Montréal, Canada. He had his Master's degree in Computer Science from the Lebanese American University (LAU) in Beirut. His main research interests are in the areas of Internet of Things, Federated Learning and Game Theory.
\end{IEEEbiography}

\vskip 0pt plus -1fil

\begin{IEEEbiography}[{\includegraphics[width=1in,height=1.25in,clip,keepaspectratio]{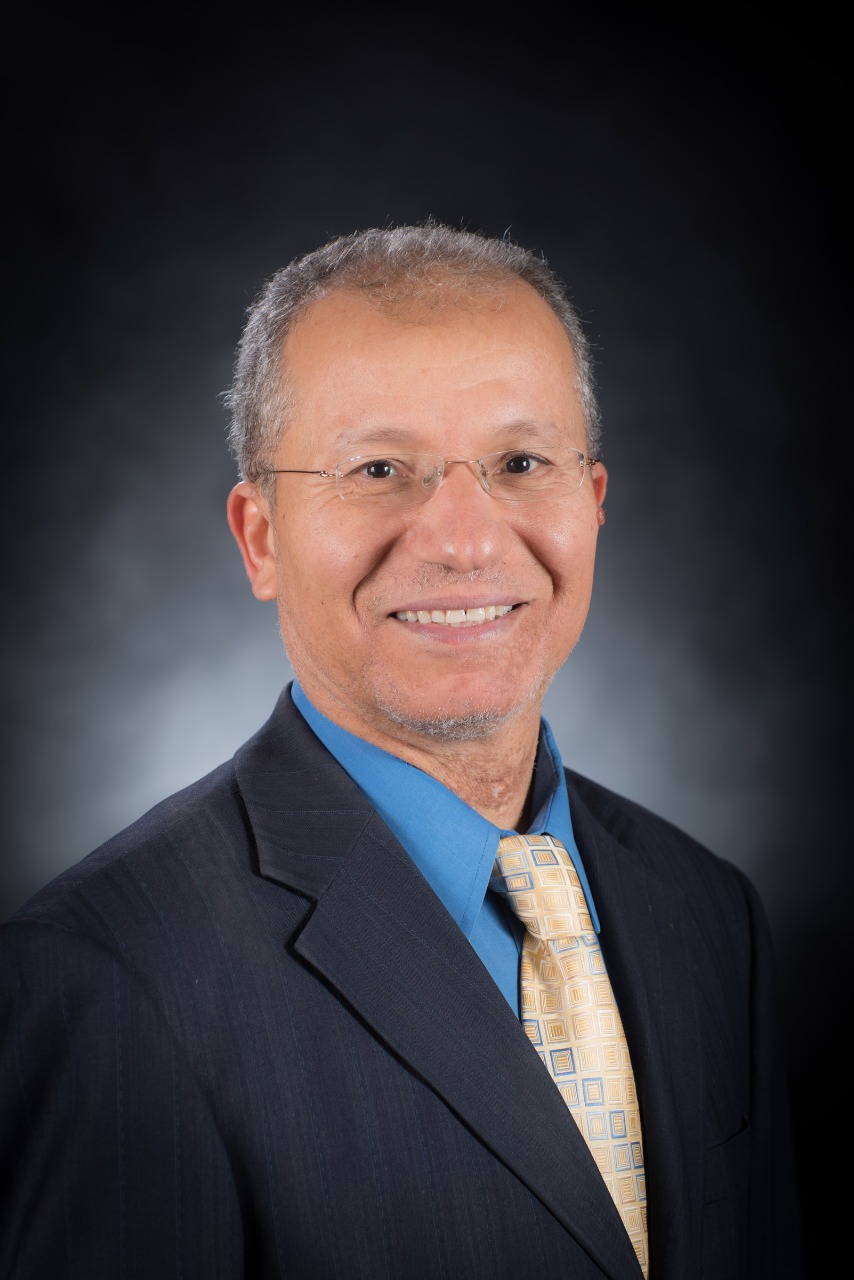}}]%
{\textbf{\textit{\textbf{Mohsen Guizani}}}}
(Fellow, IEEE) received the BS (with distinction), MS and Ph.D. degrees in Electrical and Computer engineering from Syracuse University, Syracuse, NY, USA, in 1985, 1987 and 1990, respectively. He is currently a Professor of Machine Learning and the Associate Provost at Mohamed Bin Zayed University of Artificial Intelligence (MBZUAI), Abu Dhabi, UAE. Previously, he worked in different institutions in the USA. His research interests include applied machine learning and artificial intelligence, the Internet of Things (IoT), intelligent autonomous systems, smart city, and cybersecurity. He was elevated to the IEEE Fellow in 2009 and was listed as a Clarivate Analytics Highly Cited Researcher in Computer Science in 2019, 2020 and 2021. Dr. Guizani has won several research awards, including the “2015 IEEE Communications Society Best Survey Paper Award”, the Best ComSoc Journal Paper Award in 2021 as well five Best Paper Awards from ICC and Globecom Conferences. He is the author of ten books and more than 800 publications. He is also the recipient of the 2017 IEEE Communications Society Wireless Technical Committee (WTC) Recognition Award, the 2018 AdHoc Technical Committee Recognition Award, and the 2019 IEEE Communications and Information Security Technical Recognition (CISTC) Award. He served as the Editor-in-Chief of the IEEE Network and is currently serving on the Editorial Boards of many IEEE Transactions and Magazines. He was the Chair of the IEEE Communications Society Wireless Technical Committee and the Chair of the TAOS Technical Committee. He served as the IEEE Computer Society Distinguished Speaker and is currently the IEEE ComSoc Distinguished Lecturer. 
\end{IEEEbiography}

\vskip 0pt plus -1fil

\begin{IEEEbiography}[{\includegraphics[width=1in,height=1.25in,clip,keepaspectratio]{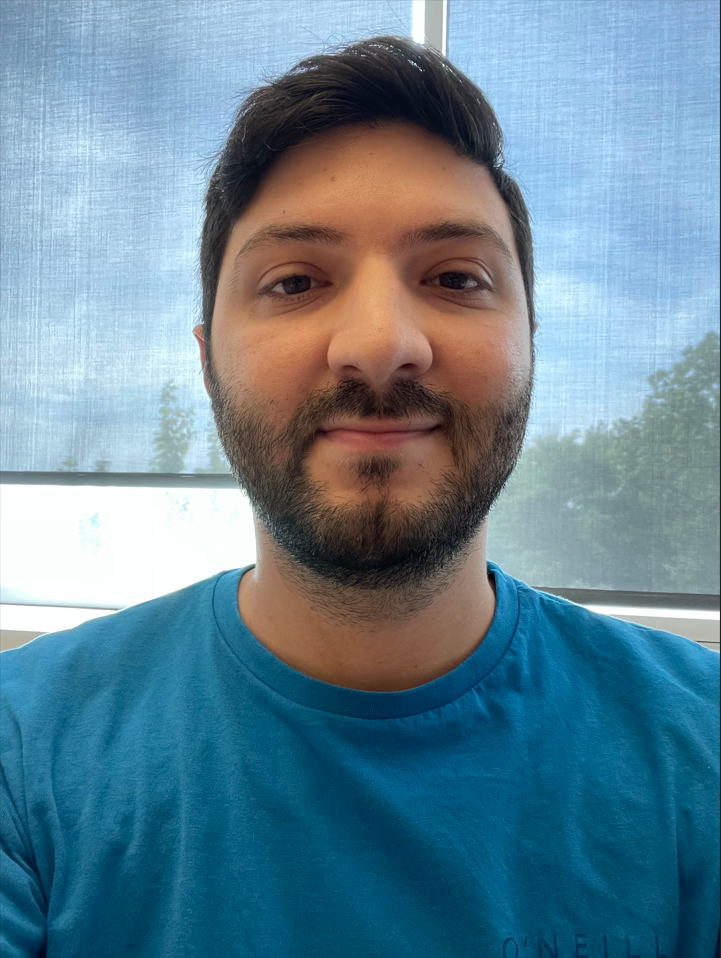}}]%
{\textbf{\textit{\textbf{Omar Abdel Wahab}}}}
received the M.Sc. degree in computer science from Lebanese American University, Beirut, Lebanon, in 2013, and the Ph.D. degree in information and systems engineering from Concordia University, Montreal, QC, Canada. He is an Assistant Professor at the Department of Computer and Software Engineering, Polytechnique Montréal, Canada. From January 2019 till July 2022, he was Assistant Professor at Université du Québec en Outaouais, Gatineau, QC, Canada. In 2017, he did a Postdoctoral Fellowship with the École de Technologie Supérieure, Montreal, where he worked on an industrial research project in collaboration with Rogers and Ericsson. His current research activities are in the areas of cybersecurity, Internet of Things and artificial intelligence. Dr. Wahab is a recipient of many prestigious grants from prestigious agencies in Canada, such as the Natural Sciences and Engineering Research Council of Canada and Mitacs. He is a TPC member and a publicity chair of several prestigious conferences, and a reviewer of several highly-ranked journals.
\end{IEEEbiography}
\vskip 0pt plus -1fil

\begin{IEEEbiography}[{\includegraphics[width=1in,height=1.25in,clip,keepaspectratio]{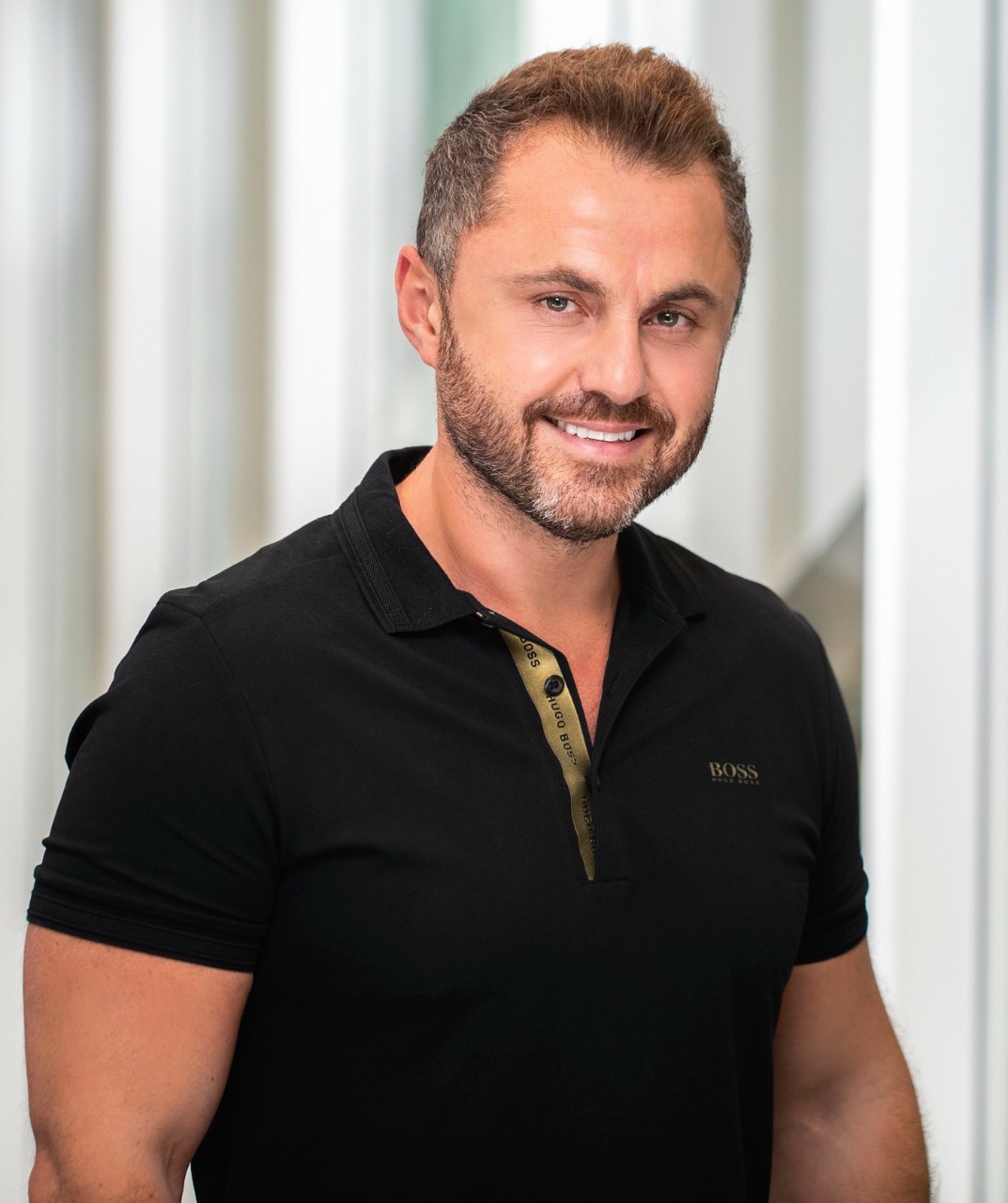}}]%
{\textbf{\textit{\textbf{Azzam Mourad}}}}is currently a Visiting Professor at Khalifa University, a Professor of Computer Science and Founding Director of the Artificial Intelligence and Cyber Systems Research Center at the Lebanese American University, and an Affiliate Professor at the Software Engineering and IT Department, Ecole de Technologie Superieure (ETS), Montreal, Canada. He was a Visiting Professor at New York University Abu Dhabi. His research interests include Cyber Security, Federated Machine Learning, Network and Service Optimization and Management targeting IoT and IoV, Cloud/Fog/Edge Computing, and Vehicular and Mobile Networks. He has served/serves as an associate editor for IEEE Transactions on Services Computing, IEEE Transactions on Network and Service Management, IEEE Network, IEEE Open Journal of the Communications Society, IET Quantum Communication, and IEEE Communications Letters, the General Chair of IWCMC2020-2022, the General Co-Chair of WiMob2016, and the Track Chair, a TPC member, and a reviewer for several prestigious journals and conferences. He is an IEEE senior member.
\end{IEEEbiography}

\vskip 0pt plus -1fil

\begin{IEEEbiography}[{\includegraphics[width=1in,height=1.25in,clip,keepaspectratio]{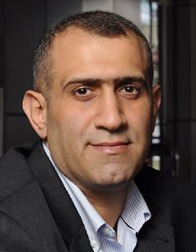}}]%
{\textbf{\textit{\textbf{Hadi Otrok}}}}
(senior member, IEEE) received his Ph.D. in ECE from Concordia University, Montreal, QC, Canada, in 2008.He holds a Full Professor position in the Department of Computer Science at Khalifa University, Abu Dhabi, UAE. He is also an Affiliate Associate Professor in the Concordia Institute for Information Systems Engineering at Concordia University, and an Affiliate Associate Professor in the Electrical Department at Ecole de Technologie Superieure (ETS), Montreal, Canada. His research interests include the domain of blockchain, reinforcement learning, federated learning, crowd sensing and sourcing, ad hoc networks, and cloud security. He co-chaired several committees at various IEEE conferences. He is also an Associate Editor at IEEE Transactions on Network and Service Management (TNSM), IEEE Transactions on Service Computing, and Ad-hoc networks (Elsevier). He also served in the editorial board of IEEE Networks and IEEE Communication Letters.
\end{IEEEbiography}
\vskip 0pt plus -1fil

\begin{IEEEbiography}[{\includegraphics[width=1in,height=1.25in,clip,keepaspectratio]{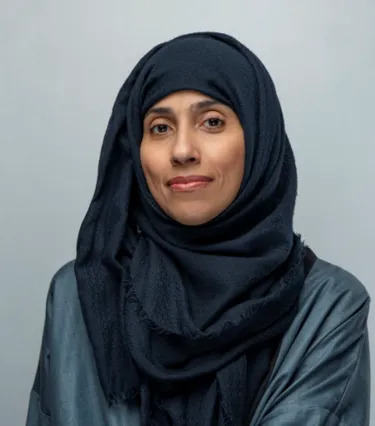}}]%
{\textbf{\textit{\textbf{Hoda Al Khzaimi}}}}
is currently the Director of the Center of Cyber Security at New York University AD and a research assistant professor at New York University. She served in different posts for research and development in the Technology development sector for the past years; she also played a vital role in developing Cyber Security and Cryptology research and development ecosystems. She headed the Department of Research and Development for Cyber Security and Cryptology in different national initiatives in the United Arab Emirates, along with her associations with different security initiatives nationally and internationally. She consults for special projects with national and international technology development initiatives.
\end{IEEEbiography}
\vskip 0pt plus -1fil

\begin{IEEEbiography}[{\includegraphics[width=1in,height=1.25in,clip,keepaspectratio]{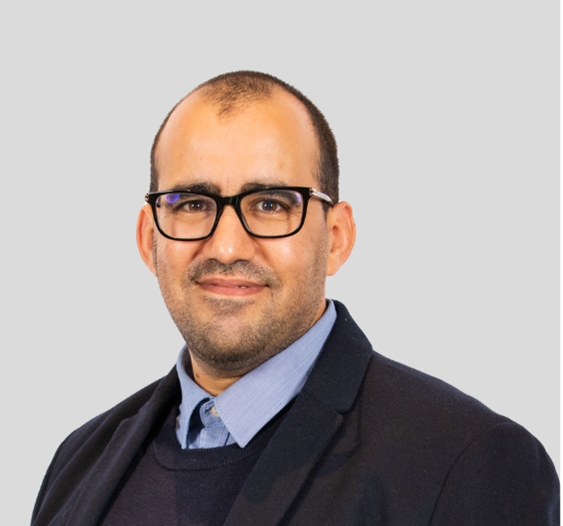}}]%
{\textbf{\textit{\textbf{Bassem Ouni}}}}
(Senior Member, IEEE) received the Ph.D. degree in computer science from the University of Nice-Sophia Antipolis, Nice, France, in July 2013. Between October 2018 and January 2022, he was a Lead Researcher with the French Atomic Energy Commission (CEA), LIST Institute, Paris, France, and an Associate Professor/a Lecturer with the University of Paris Saclay and ESME Sudria Engineering School, Paris. Prior to that, he was a Lead Researcher with the Department of Electronics and Computer Science, University of Southampton, Southampton, U.K., between 2017 and 2018. Before that, he was a Research Scientist with the Institute of Technology in Aeronautics, Space and Embedded Systems (IRT-AESE), Toulouse, France, between 2015 and 2016. From September 2013 to 2014, he held a postdoctoral fellow position with EURECOM, Telecom ParisTech Institute, and Sophia Antipolis, France. Furthermore, he was a Lecturer with the University of Nice Sophia Antipolis (Polytech Nice Engineering School and Faculty of Sciences of Nice), between 2009 and 2013. He is currently a Lead Researcher with the Technology Innovation Institute, Abu Dhabi, United Arab Emirates. Also, he was managing several industrial collaborations with ARM, Airbus Group Innovation, Rolls Royce, Thales Group, Continental, and ACTIA Automotive Group. He has coauthored many publications (book chapters, journals, and international conferences).
\end{IEEEbiography}

\end{document}